\documentclass{aastex63}
\usepackage{graphicx}
\usepackage{subfigure}
\usepackage{multirow}
\usepackage{gensymb}

\def\na{Nature}
\def\nat{Nature}
\def\apj{ApJ}
\def\aj{AJ}
\def\apjl{ApJ}
\def\apjs{ApJ}

\def\araa{Ann. Rev. Astr. Ap}
\def\mnras{MNRAS}

\def\aap{A\&A}

\def\araa{ARA\&A}

\shorttitle{Pulsar in NGC~6712}
\shortauthors{Yan et al.}

\begin{document}

\title{An Eclipsing Black Widow Pulsar in NGC~6712}

\correspondingauthor{Zhen Yan}
\email{yanzhen@shao.ac.cn}

\author{Zhen Yan}
\affiliation{Shanghai Astronomical Observatory, Chinese Academy of Sciences, NO.~80 Nandan Road, Shanghai 200030, China}
\affiliation{University of Chinese Academy of Sciences, No.19A Yuquan Road, Shijingshan District, Beijing 100049, China}
\affiliation{Key Laboratory of Radio Astronomy, Chinese Academy of Sciences£¬10 Yuanhua Road, Nanjing, JiangSu 210033, China}

\author{Zhi-chen Pan}
\affiliation{National Astronomical Observatories, Chinese Academy of Sciences, 20A Datun Road, Chaoyang District, Beijing 100101, China}
\affiliation{CAS Key Laboratory of FAST, National Astronomical Observatories, Chinese Academy of Sciences, Beijing 100101, China}

\author{Scott M. Ransom}
\affiliation{National Radio Astronomy Observatory, Charlottesville, VA 22903, USA}

\author{Duncan R. Lorimer}
\affiliation{Department of Physics and Astronomy, West Virginia University, Morgantown, WV 26506, USA}
\affiliation{Center for Gravitational Waves and Cosmology, West Virginia University, Chestnut Ridge Research Building, Morgantown, WV 26505, USA}

\author{Lei Qian}
\affiliation{National Astronomical Observatories, Chinese Academy of Sciences, 20A Datun Road, Chaoyang District, Beijing 100101, China}
\affiliation{CAS Key Laboratory of FAST, National Astronomical Observatories, Chinese Academy of Sciences, Beijing 100101, China}

\author{Pei Wang}
\affiliation{National Astronomical Observatories, Chinese Academy of Sciences, 20A Datun Road, Chaoyang District, Beijing 100101, China}
\affiliation{CAS Key Laboratory of FAST, National Astronomical Observatories, Chinese Academy of Sciences, Beijing 100101, China}

\author{Zhi-qiang Shen}
\affiliation{Shanghai Astronomical Observatory, Chinese Academy of Sciences, NO.~80 Nandan Road, Shanghai 200030, China}
\affiliation{University of Chinese Academy of Sciences, No.19A Yuquan Road, Shijingshan District, Beijing 100049, China}
\affiliation{Key Laboratory of Radio Astronomy, Chinese Academy of Sciences£¬10 Yuanhua Road, Nanjing, JiangSu 210033, China}

\author{Di Li}
\affiliation{University of Chinese Academy of Sciences, No.19A Yuquan Road, Shijingshan District, Beijing 100049, China}
\affiliation{National Astronomical Observatories, Chinese Academy of Sciences, 20A Datun Road, Chaoyang District, Beijing 100101, China}
\affiliation{CAS Key Laboratory of FAST, National Astronomical Observatories, Chinese Academy of Sciences, Beijing 100101, China}

\author{Peng Jiang}
\affiliation{University of Chinese Academy of Sciences, No.19A Yuquan Road, Shijingshan District, Beijing 100049, China}
\affiliation{National Astronomical Observatories, Chinese Academy of Sciences, 20A Datun Road, Chaoyang District, Beijing 100101, China}
\affiliation{CAS Key Laboratory of FAST, National Astronomical Observatories, Chinese Academy of Sciences, Beijing 100101, China}

\author{Jin-Tao Luo}
\affiliation{University of Chinese Academy of Sciences, No.19A Yuquan Road, Shijingshan District, Beijing 100049, China}
\affiliation{National Time Service Center, Chinese Academy of Sciences, Xi¡¯an 710600, China}

\author{Jie Liu}
\affiliation{Shanghai Astronomical Observatory, Chinese Academy of Sciences, NO.~80 Nandan Road, Shanghai 200030, China}
\affiliation{University of Chinese Academy of Sciences, No.19A Yuquan Road, Shijingshan District, Beijing 100049, China}

\author{Zhi-peng Huang}
\affiliation{Shanghai Astronomical Observatory, Chinese Academy of Sciences, NO.~80 Nandan Road, Shanghai 200030, China}
\affiliation{University of Chinese Academy of Sciences, No.19A Yuquan Road, Shijingshan District, Beijing 100049, China}

\begin{abstract}
We report the discovery of the first radio pulsar associated with NGC~6712, an eclipsing black widow (BW) pulsar, J1853$-$0842A, found by high-sensitivity searches using the Five-hundred-meter Aperture Spherical radio Telescope. This 2.15~ms pulsar is in a 3.56~hr compact circular orbit with a very low mass companion likely of mass 0.018 to 0.036~$M_{\rm \odot}$ and exhibits eclipsing of the pulsar signal. Though the distance to PSR~J1853$-$0842A predicted from its dispersion measure ($155.125 \pm 0.004$~cm$^{-3}$~pc) and Galactic free electron density models are about 30\% smaller than that of NGC~6712 obtained from interstellar reddening measurements, this is likely due to limited knowledge about the spiral arms and Scutum stellar cloud in this direction. Follow-up timing observations spanning 445 days allow us to localize the pulsar's position to be 0.14 core radii from the center of NGC~6712 and measure a negative spin-down rate for this pulsar of $-2.39(2)\times10^{-21}\rm~s~s^{-1}$. The latter cannot be explained without the acceleration of the GC and decisively supports the association between PSR~J1853--0842A and NGC~6712. Considering the maximum GC acceleration, Galactic acceleration, and Shklovskii effect, we place an upper limit on the intrinsic spin-down rate to be $1.11\times10^{-20}\rm~s~s^{-1}$. From an analysis of the eclipsing observations, we estimate the electron density of the eclipse region to be about $1.88\times10^6\rm~cm^{-3}$. We also place an upper limit of the accretion rate from the companion is about $3.05\times10^{-13}~M_{\rm \odot}\rm~yr^{-1}$ which is comparable with some other BWs.
\end{abstract}


\section{Introduction} \label{sec:intro}
Globular Clusters (GCs) are tightly bound by gravity, giving them spherical shapes and relatively high stellar densities toward their centers. This environment provides a high specific incidence of low-mass X-ray binaries (LMXBs), the proposed progenitors of millisecond pulsars (MSPs), which enhances the possibility of finding new pulsars in GCs compared to the Galactic disk \citep[see, e.g,][]{pla+03, ran08}. Since the discovery of first pulsar in M~28 \citep{lbm+87}, there are currently about 217 pulsars known to be associated with Galactic GCs; most of these are binary MSPs \citep{fre13}\footnote{\url{http://www3.mpifr-bonn.mpg.de/staff/pfreire/GCpsr.html}}. In the core of a GC, more frequent interactions between stars take place as the stellar densities get even higher. The stellar interactions also produce exotic MSP binaries which are extremely rare in other places. The MSP is thought to have been spun up by the transfer of matter and angular momentum from its low-mass companion star during an X-ray-emitting phase. When a neutron star has been spun-up to millisecond periods, its strong radiation has a possibility to quench accretion by ablating surrounding plasma, potentially evaporating the companion entirely to form an isolate MSP \citep{rst89}. Support for this scenario is provided by discoveries of eclipsing redback (RB) \citep{asr09} and black widow (BW) pulsars \citep{fst88}. These systems typically have compact orbits with periods shorter than 24~hr and are accompanied by low-mass companions with typical masses a few tenths of a solar mass or less ($<0.1~M_{\rm \odot}$ for BWs and $0.2-0.4~M_{\rm \odot}$ for RBs). Evolutionary studies of BWs and RBs can provide important links between LMXBs and MSPs \citep{rob13}.

According to the catalog of GCs in the Milky Way, NGC~6712 is a metal-rich GC located about 6.9~kpc away from the Sun. The core and half-light radii of NGC~6712 are 0.76 and 1.33 arcmin, respectively. In Table \ref{table:ngc6712}, we list useful properties of NGC~6712 relate to this work obtained from the catalog\footnote{\url{http://physwww.mcmaster.ca/~harris/mwgc.dat}} of GCs \citep{har96}. The stellar encounter rate for a GC can be estimated with $\Gamma \propto \rho_{\rm c}^{1.5} r_{\rm c}^2$, where $\rho_{\rm c}$ and $r_{\rm c}$ is the density and radius of the cluster core, respectively \citep{vh87}. For NGC~6712, $\Gamma$ is less than $1\%$ of Terzan~5, where 39 pulsars have been discovered. In NGC~6712, there is a LMXB with the orbital period ($P_{\rm b}$) of 0.33~hr detected with the Einstein Observatory's Monitor Proportional Counter \citep{lde+83}. In spite of a number of radio pulsar searches in NGC~6712 previously carried out, prior to this work, no pulsars were known in this cluster. This led to upper limits on pulsed emission of 11~$\rm mJy$ at 400~MHz \citep{bl96} and 16~$\rm \mu Jy$ at 2.0~GHz \citep{lrf11}. Using scaling laws derived in previous population studies, which show that $\Gamma$ is a strong indicator of pulsar abundance in GCs \citep{hct10,tl13}, we estimate the population of pulsars in NGC~6712 to be $\sim 5$ --- much smaller than Terzan~5 and highlighting the importance of sensitive surveys such as the work described in this paper.

\begin{table}
\centering
\caption{Observed properties of NGC~6712}
\label{table:ngc6712}
\begin{tabular}{l c}
\hline\hline
Name & NGC~6712 \\
Right Ascension (J2000)    &    $18^{\rm h}53^{\rm m}04.3^{\rm s}$ \\
Declination (J2000)   &   $-08^{\circ}42'22.0''$   \\
Galactic longitude, $l$, (deg) &  25.35  \\
Galactic latitude, $b$, (deg) &   -4.32 \\
Distance, $D$, (kpc) & 6.9 \\
Core radius, $r_{\rm c}$, (arcmin) & 0.76 \\
Half-light radius, $r_{\rm h}$, (arcmin) &  1.33 \\
Central velocity dispersion, $\sigma_{\rm v}$  (km/s) & 4.3 \\
\hline
\end{tabular}
\end{table}

The Five-hundred-meter Aperture Spherical radio Telescope (FAST) is the largest single dish radio telescope in the world \citep{nan06}. Benefiting from its 300-m illuminated aperture and low-noise cryogenic receivers, FAST can perform unprecedented high-sensitivity observations \citep{jyg19}. In this paper, we will present pulsar search results of NGC~6712 with FAST which have led to the discovery of the first pulsar in this cluster, PSR~J1853--0842A. The rest of this paper is structured as follows: in Section 2, we describe the observations and data reduction procedures; in Section 3, we present the results of the search and follow-up timing observations; in Section 4, we discuss the implications and significance of our results.

\section{Observations and Data Reduction}
Observations of NGC~6712 were carried out with FAST using the central beam of the 19-beam L-band receiver \citep{jth20}. For these observations, where the typical system temperature is about 20~K, the unparalleled gain of FAST leads to a system equivalent flux density of about 2~Jy, at least a factor of three improvement over the previous observations mentioned above. To carry out a dispersion measure (DM)
search, the total bandwidth (1.05$-$1.45~GHz) was divided into sub-channels of 0.12~MHz and the data were acquired using the pulsar searching mode with a time resolution of 49.152~$\rm \mu s$. Following our discovery of the pulsar in the initial 30-min observation carried out on 2019 June 25 (MJD~58659), confirmation and follow-up observations were arranged on MJDs~58685 (observation length $T_{\rm o} = 120$~min, $T_{\rm o}$ will be omitted hereafter), 58686 (60~min), 58687 (30~min), 58768 (30~min), 58769 (30~min), 58931 (60~min), 58933 (60~min), 58963 (10~min), 58965 (10~min), and 59105 (30~min).

All data were searched for the presence of periodic dispersed pulses using
\texttt{PRESTO}\footnote{\url{https://www.cv.nrao.edu/~sransom/presto}} (PulsaR Exploration and Search TOolkit) \citep{ran11}. Within \texttt{PRESTO}, the routine \texttt{rfifind} was used to mask and zap radio-frequency interference in both the time and frequency domains. The predicted DM of NGC~6712 is about 182~$\rm cm^{-3}~pc$ based on its distance (6.9~kpc) \citep{har96} and the YMW16 Galactic free electron density models \citep{ymw17}. As no pulsar was discovered in NGC~6712 previously, we searched for periodic signals in a DM range 0--300~$\rm pc~cm^{-3}$ with the step of 0.05~$\rm pc~cm^{-3}$. The \texttt{PRESTO} routine \texttt{accelsearch} was used to analyse the data using a Fourier-domain acceleration search technique \citep{rem02} which, for a spin frequency $f$ and an observation length $T$, was sensitive to frequency drifts $z=\dot{f}T^2$ of up to $z_{\rm max} = 200$~Fourier bins. To search for even more highly accelerated binary systems, we also used the so-called ``jerk search'' technique \citep{ar18} which looks for signal drifts $w=\dot{z}=\ddot{f}T^3$. In our searches, we set the maximum drift $w_{\rm max}=600 $ bins. The python script \texttt{ACCEL\_sift.py} was used to produce a winnowed list of pulsar candidates from the searches. Beside the periodic pulse signal, we also used \texttt{single\_pulse\_search.py} to search for single pulses by setting the threshold to be 5.0.

\section{Results}
We found a promising pulsar candidate with a spin period $P=2.149$~ms and DM of 155.13~$\rm cm^{-3}~pc$ in the initial observation from MJD~58659. With jerk search, the signal-to-noise ratios (SNR) of candidate detection was enhanced from 44.06 to 45.97. The signal was subsequently re-detected in follow-up observations with a SNR in the range 20.26---48.39. Each detection of the pulsar was further refined using the \texttt{PRESTO} routine \texttt{prepfold} which searched the data in period and DM to produce integrated pulse profiles with sub-integration lengths of 1.2~min. These period searches provided $P$ and $\dot{P}$ measurements for each observations which were then used to obtain a preliminary estimate of the orbital parameters of the pulsar using the analysis technique described by \citet{fkl01}. This initial spin and orbital ephemeris was subsequently supplied as input to the \texttt{TEMPO}\footnote{\url{http://tempo.sourceforge.net}} software package \citep{nds15} which we used to carry out a full phase-coherent timing analysis of the time-of-arrival (TOA) for each folded profile. We used the \texttt{get\_toa.py} routine in \texttt{PRESTO} \citep[which is an implementation of Fourier-domain template matching;][]{tay92} to obtain TOAs which were analyzed in \texttt{TEMPO} using well-established methods \citep{lok04}.

\begin{figure}[htbp]
\centering
\subfigure{
\begin{minipage}{0.49\textwidth}
\includegraphics[width=0.485\textwidth,angle=-90]{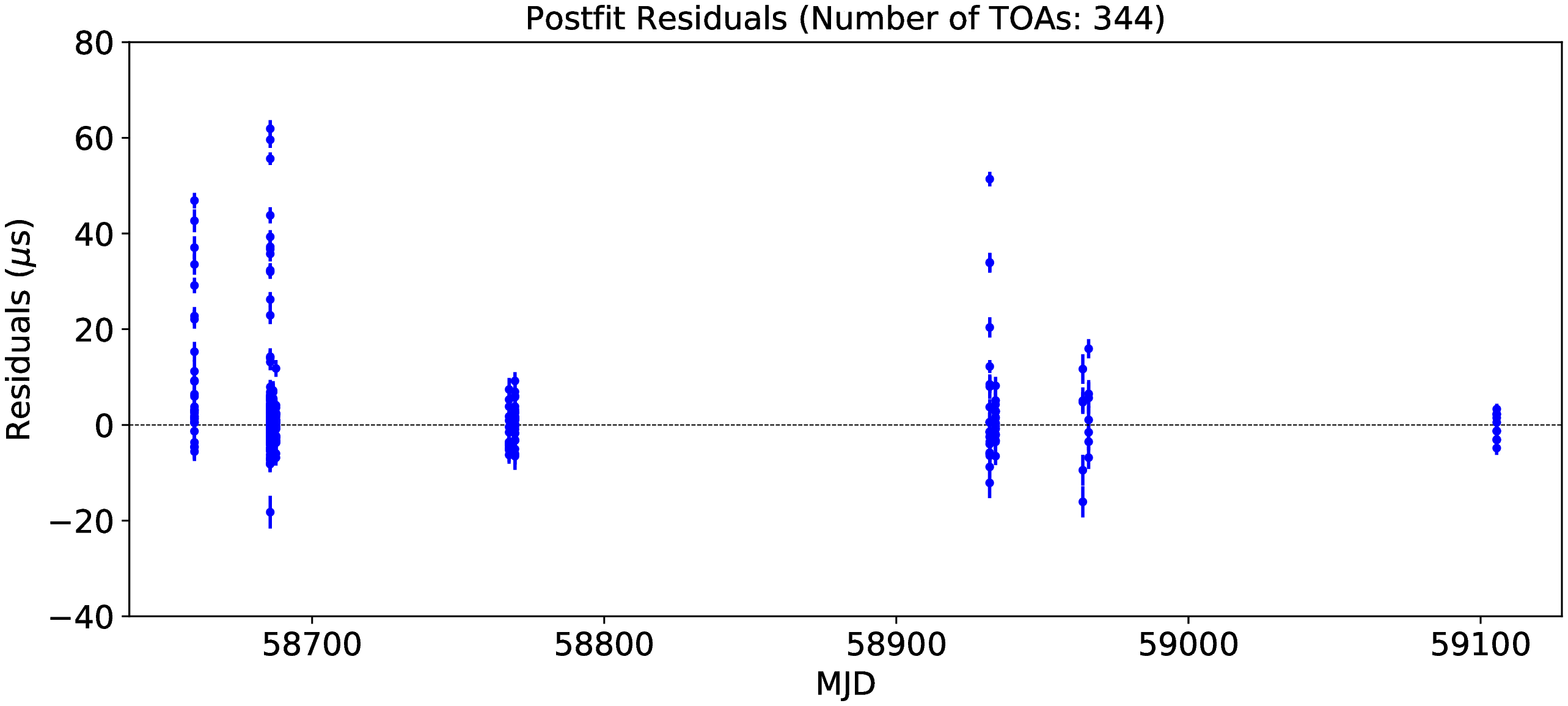}
\vspace{-0.18cm}
\end{minipage}
}
\subfigure{
\begin{minipage}{0.49\textwidth}
\includegraphics[width=0.485\textwidth,angle=-90]{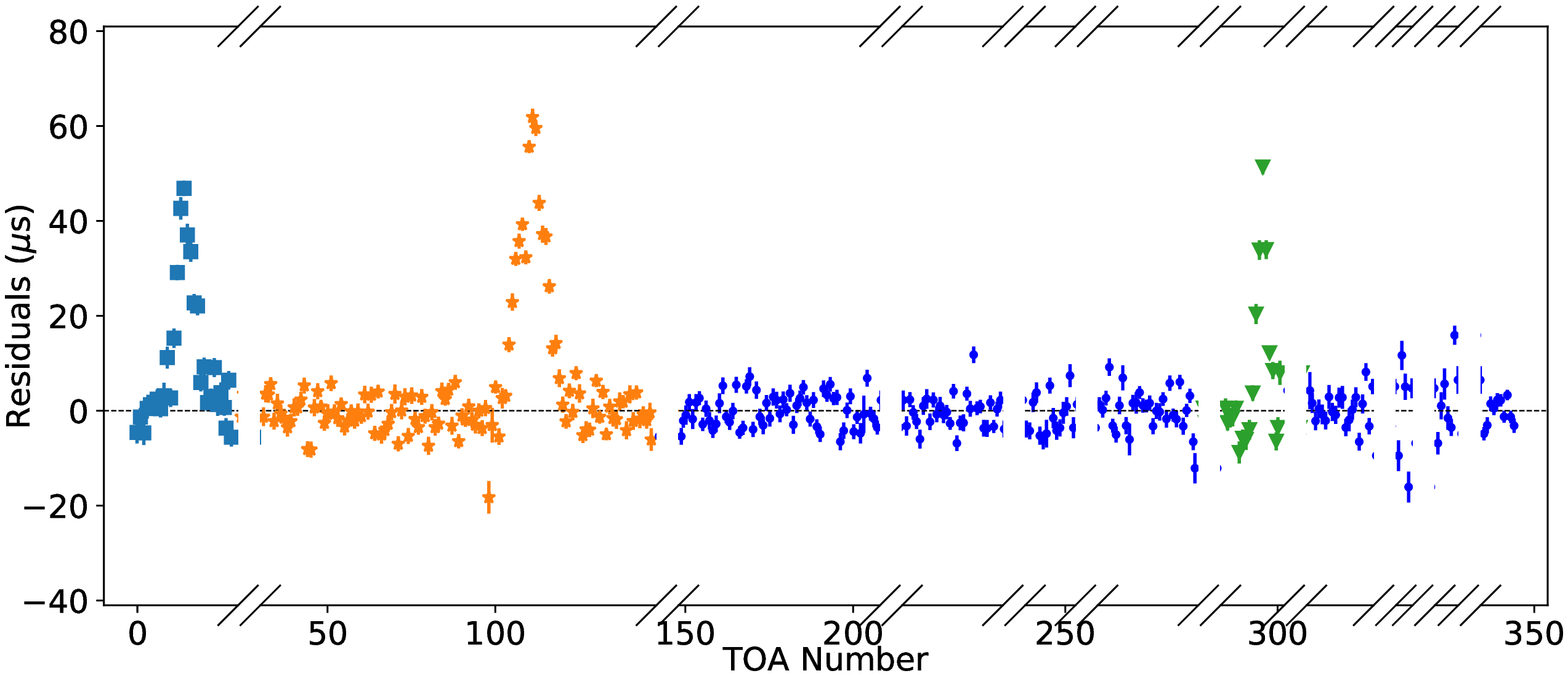}
\vspace{-0.18cm}
\end{minipage}
}
\subfigure{
\begin{minipage}{0.49\textwidth}
\includegraphics[width=0.485\textwidth,angle=-90]{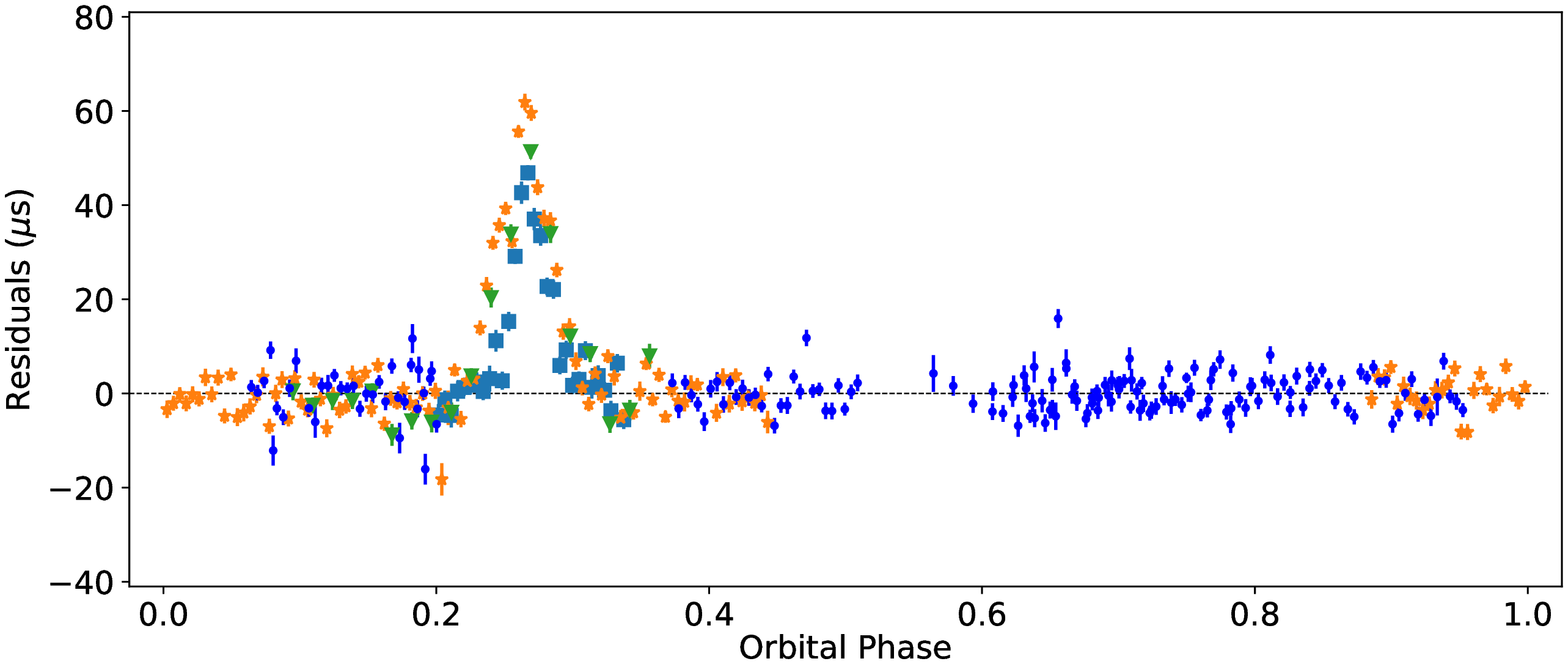}
\vspace{-0.1cm}
\end{minipage}
}
\caption{Timing residuals for PSR~J1853$-$0842A as a function of MJD (top), time-of-arrival (TOA) number (middle), and orbital phase (bottom). The cyan squares, orange stars and green triangles in the plot shown in the middle and bottom panels are corresponding to the observation results on MJD~58659, 58685 and 58931, respectively. Pulsar eclipses were detected on each of these days. In the middle panel, observation results from different epochs are separated by breaks on the horizontal axes of the plot. The integration length of each point is about 1.2~min.}
\label{fig:toaplot}
\end{figure}

The timing analysis resulted in a phase-coherent solution for the new pulsar, which we henceforth refer to as PSR~J1853$-$0842A, over the MJD range 58659--59105. The timing residuals as a function of MJD, TOA number and orbital phase are presented in Fig.~\ref{fig:toaplot}. The measured and derived parameters of PSR~J1853$-$0842A are listed in the Table~\ref{table:timingp}. As seen in Fig.~\ref{fig:toaplot}, the extra time delays take place on three separate days, all of which correspond to the same orbital phase range (0.22--0.32). These phenomena should be caused by the eclipse of ionized material surrounding the companion star, as dispersion time delay was detected if we divided the total bandwidth into two sub-bands with the central frequency of 1.15 and 1.35~GHz respectively. The emission from the MSP is ablating its companion in such a narrow binary orbit. The expected geometry of this system should the eclipsing material be spherically symmetric and centered in the orbital plane at the distance of the companion. The maximum extra time delay is detected when the companion passes closest to our line of sight towards the pulsar \citep{pbs19}.

\begin{figure}
\centering
\includegraphics[width=0.45\textwidth,angle=-90]{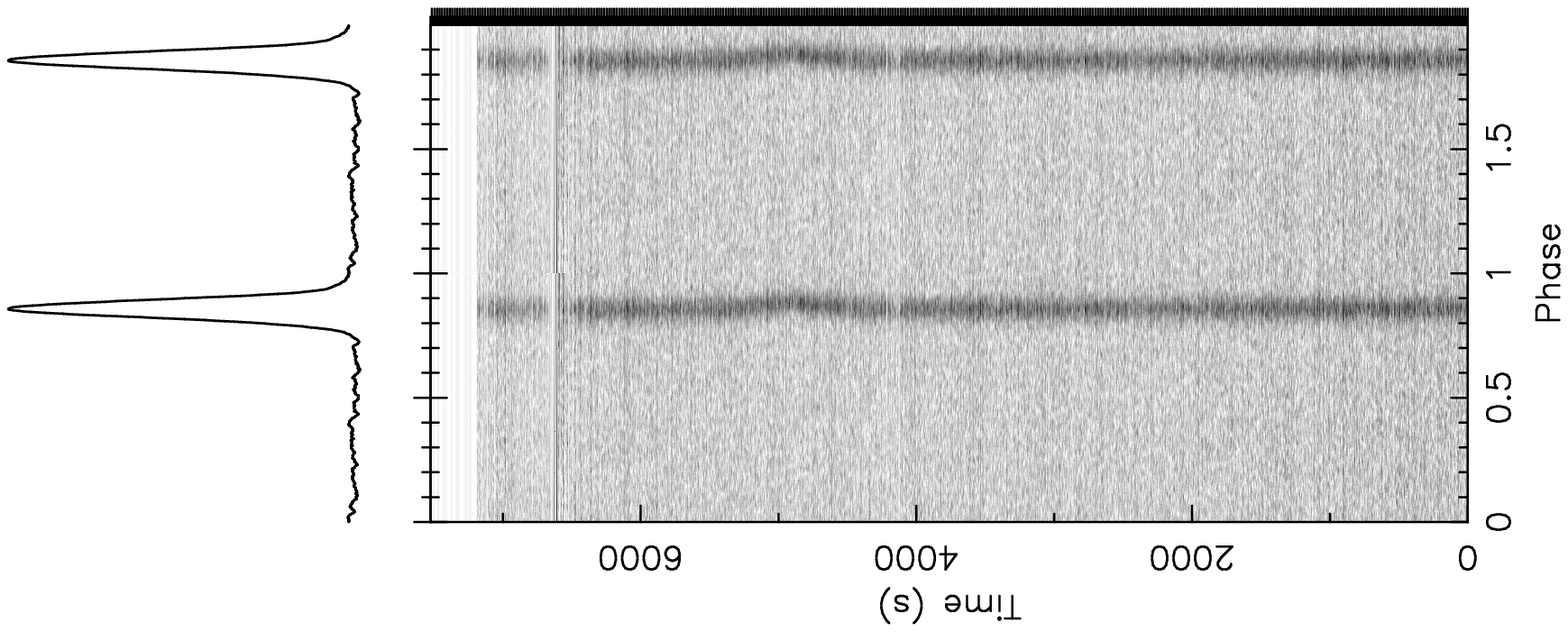}
\hspace{1.0cm}
\includegraphics[width=0.45\textwidth,angle=-90]{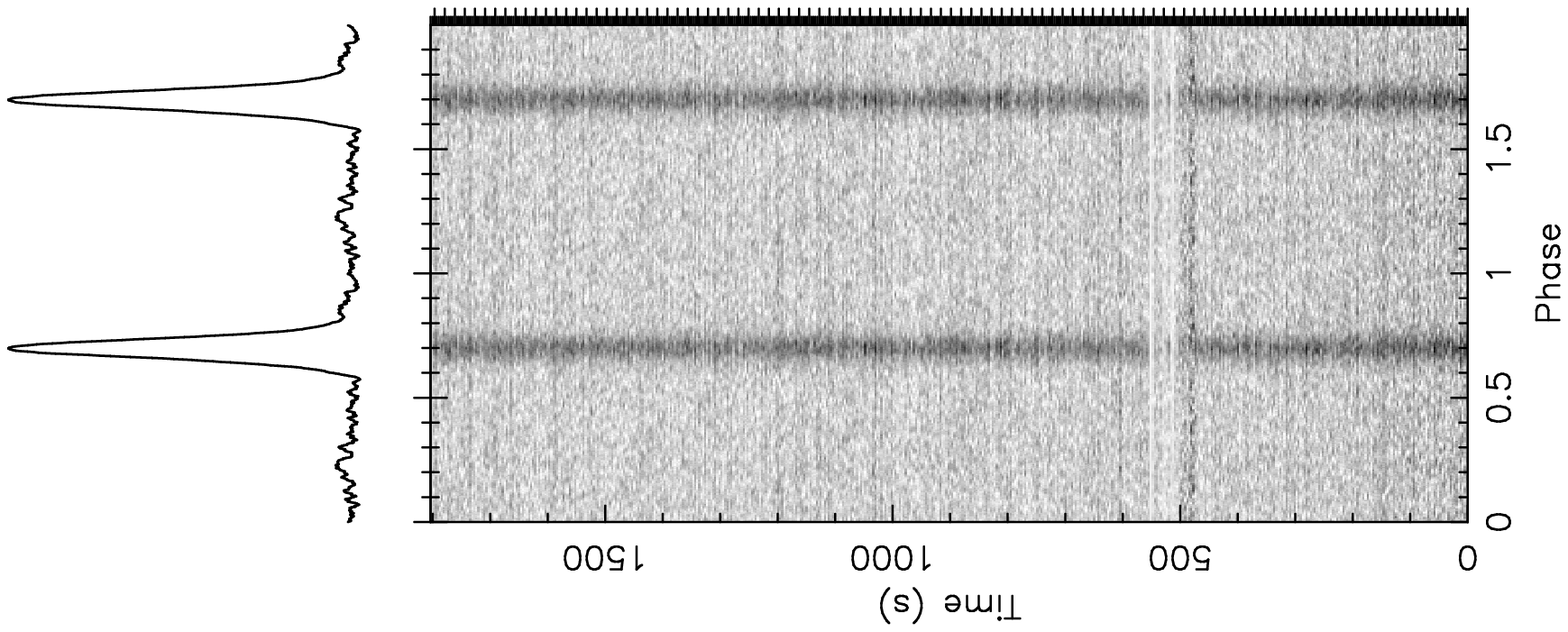}
\vspace{1.2cm}
\caption{Phase-time (bottom) and integrated profile (top) plots for PSR~J1853$-$0842A obtained with observation data on MJD~58685 (left) and 58768 (right).}
\label{fig:profileplot}
\end{figure}

In Fig.~\ref{fig:profileplot}, we present two samples of phase-time and integrated profile plots for PSR~J1853$-$0842A based on the observations from MJD~58685 and 58768. The sub-integration time for the phase-time plots is about 7~min. Obvious time delays caused by eclipse are detected around 4900~s in the phase-time plot of observation on MJD~58685. The integrated profile of PSR~J1853$-$0842A shows a simple structure with only a single peak. The widths at 50\% and 10\% of the peak of the integrated profile are $W_{\rm 50}=26.04\pm1.24^{\circ}$ and $W_{\rm 10}=63.41\pm2.56^{\circ}$, respectively.

Since there were no radio flux density calibrators arranged in our observations, we estimate the flux density of PSR~J1853$-$0842A using radiometer noise calculations \citep[see, e.g.,][]{lrf11} which gives the off-pulse root-mean-square noise,
\begin{equation}
\sigma=\frac{T_{\rm sys}}{G\sqrt{N_{\rm p} \Delta \nu T_{\rm o}}}.
\end{equation}
Here $T_{\rm sys}$, $G$, $N_{\rm p}$, $\Delta \nu $, and $T_{\rm o}$ are the system noise temperature, antenna gain, number of polarization, bandwidth and length of observation, respectively. Using the nominal values for these parameters, and adopting $G=11.3~\rm K~Jy^{-1}$ \citep{jyg19}, since the zenith angle is large in our observations, we find the estimated mean flux density of PSR~J1853$-$0842A to be $16.1\pm2.9$~$\rm \mu Jy$ at 1.25~GHz.

\begin{table}
\begin{center}
\caption{Observed and derived parameters of PSR~J1853$-$0842A}
\label{table:timingp}
\begin{tabular}{l c}
\hline\hline
Pulsar  &   J1853$-$0842A                                                             \\
Right Ascension, $\alpha$ (J2000)             & $18^{\rm h}53^{\rm m}04.07409(2)^{\rm s}$     \\
Declination, $\delta$ (J2000)                 & $-08^{\circ}42'28.254(2)''$                                     \\
Spin Frequency, $f_{\rm 0}$ (s$^{-1}$)        &   465.23897161363(7)                                             \\
$1^{\rm st}$ Spin Frequency derivative, $f_{\rm 1}$ (Hz s$^{-2}$)   &   5.18(4)$\times 10^{-16}$ \\
Reference Epoch (MJD)                        &   58685.702931                                           \\
Start of Timing Data (MJD)                   &   58659.715                                            \\
End of Timing Data (MJD)                    &   59105.545                                             \\
Dispersion Measure, DM (pc cm$^{-3}$)        &   155.125(4)                                          \\
Solar System Ephemeris       &   DE200                                                                  \\
Number of TOAs           &   344                                                                    \\
Residuals RMS ($\mu$s)    &   3.77                                                                   \\
\hline
\multicolumn{2}{c}{Binary Parameters}  \\
\hline\hline
Binary Model          &   BT                                                                     \\
Projected Semi-major Axis, $x_{\rm p}$ (lt-s)   &   4.91856(2)$\times 10^{-2}$                         \\
Orbital Eccentricity, $e$    &   0.00                                                                      \\
Longitude of Periastron, $\omega$ (deg) &   0.00                                    \\
Epoch of passage at Periastron, $T_0$ (MJD)  &   58685.64161943(9)                   \\
Orbital Period, $P_b$ (days)  &   0.1482829972(2)                       \\
\hline
\multicolumn{2}{c}{Derived Parameters}  \\
\hline\hline
Spin Period, $P$ (s)     &   2.1494330032835(3)$\times 10^{-3}$                                     \\
1st Spin Period derivative, $\dot{P}$ (s s$^{-1}$)  &   $-$2.39(2)$\times 10^{-21}$          \\
Distance based on TC93 \tablenote{\cite{tc93}}, $D_{\rm TC93}$ (kpc)  &  4.55\\
Distance based on NE2001 \tablenote{\cite{cl02}}, $D_{\rm NE2001}$ (kpc) &  3.79   \\
Distance based on YMW16 \tablenote{\cite{ymw17}}, $D_{\rm YMW16}$ (kpc) &  4.76 \\
Mass function, $f(M_{\rm p}, M_{\rm c})$ ($M_{\rm \odot}$) & $5.81\times10^{-6}$\\
Possible range of $M_{\rm c}$ \footnote{For $i\geq60.0^{\circ}$ and $1.0~M_{\rm \odot} \leq M_{\rm p} \leq 2.2~M_{\rm \odot}$ (see Sect.~\ref{sec:disc})} ($M_{\rm \odot}$), & $0.018\leq M_{\rm c}\leq 0.036$  \\
\hline
\end{tabular}
\end{center}
\end{table}

\section{Discussion}
\label{sec:disc}
\begin{figure}
\centering
\includegraphics[width=0.5\textwidth,angle=0]{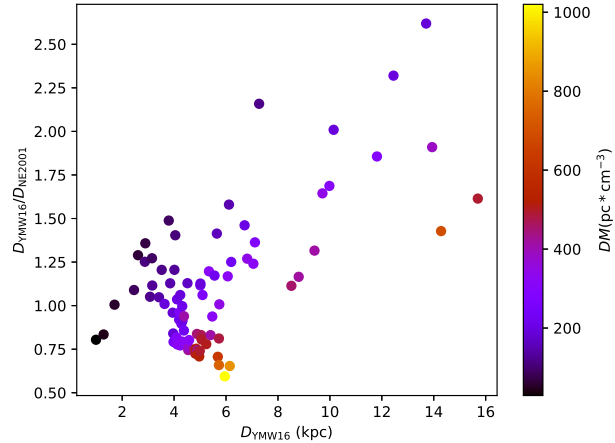}
\hspace{1.0cm}
\caption{$D_{\rm YMW16}/D_{\rm NE2001}$ changes with $D_{\rm YMW16}$ for 87 pulsars within $5^\circ$ around NGC~6712. The DM value of each pulsar is scaled with the color bar on the right.}
\label{fig:distratio}
\end{figure}

Beside the 6.9~kpc adopted in the Galactic GC catalog \citep{har96}, there are some other distance results of NGC~6712, e.g. 6.75~kpc \citep{ss+66}, 6.2~kpc \citep{web85}, 7.9 kpc \citep{omb00}, which were also obtained base on interstellar reddening measurements. Overall, DM-based distances of PSR~J1853$-$0842A listed in Table~\ref{table:ngc6712} are much smaller than previous distance results of NGC~6712. Normally, distance results obtained with interstellar reddening measurements are comparatively reliable, as they are supported by model-independent parallax measurements on a large sample of stars at optical band \citep{fuh04}. By comparison, only 144 pulsars' distances are obtained with parallax measurements\footnote{\url{http://hosting.astro.cornell.edu/~shami/psrvlb/parallax.html}}. To give further estimate on whether PSR~J1853$-$0842A is in NGC~6712 or not. We firstly did an assessment about the validation of distance based on the Galactic electron density models along this direction. Comparisons of model-independent distances of about 50 pulsars obtained with the Very Long Baseline Array (VLBA) to those predicted by the NE2001 \citep{cl02} and YMW16 \citep{ymw17} Galactic electron density distribution models show that it is hard to give definite conclusion on which model is more accurate. Both models show large errors for some objects, with the NE2001 model doing better on some objects and the YMW16 model on others. For about 14\% of those 50 pulsars, their distances based on the YMW16 model fall outside the range of 0.1 to 1.9 times of corresponding real results obtained with the VLBA \citep{dgb19}. The Galactic longitude and latitude of NGC~6712 is $25.35^\circ$ and $-4.32^\circ$, respectively. We have limited information about spiral arms and Scutum stellar cloud along this direction. There is only one pulsar that has model-independent distance measurement result within $5^\circ$ around NGC~6712. As we do not know which Galactic electron density model is more accurate, we give a statistic on $D_{\rm YMW16}$ and $D_{\rm NE2001}$ of 87 pulsars within $5^\circ$ around NGC~6712. The distances of these pulsars are obtained from pulsar catalog (PSRCAT) \citep{mht05}\footnote{\url{http://www.atnf.csiro.au/research/pulsar/psrcat}}. Fig.~\ref{fig:distratio} shows the $D_{\rm YMW16}/D_{\rm NE2001}$ ratios of these 87 pulsars change with $D_{\rm YMW16}$. It is clear the ratio gets larger as the $D_{\rm YMW16}$ gets further. When the $D_{\rm YMW16}$ is larger than 6.0~kpc, the ratio can get as large as more than 2.5 times. So, the association between PSR~J1853$-$0842A and NGC~6712 can not simply be excluded base on their large model-dependent distance differences mentioned above because of their unknown accuracy.

The rotation of a pulsar usually slows down with time, as its rotational energy is converted to radiation power. So, the intrinsic spin-down rate ($\dot P_{\rm int}$) ought to be a positive value. But, the observed spin-down rate ($\dot P_{\rm obs}$) of PSR~J1853$-$0842A is about $-2.39\times10^{-21}$~$\rm~s~s^{-1}$, which is a negative result. The $\dot P_{\rm int}$ of a pulsar (especially MSP) is usually contaminated by acceleration of the host Galaxy ($a_{\rm G}$) and the Shklovskii effect caused by its proper motion \citep{shk70}. In addition, the acceleration effect of the host GC ($a_{\rm L}$) should also be considered as it sometimes
gives {\bf the dominant} contribution to the $\dot P_{\rm obs}$ of the pulsar in the GC. According to \cite{phi93},
\begin{equation}
\dot P_{\rm obs}=\dot P_{\rm int}+\frac{a_{\rm G}}{c} P+\frac{V_{\rm \perp}^2}{cD} P+\frac{a_{\rm L}}{c} P,
\label{eq:pdot}
\end{equation}
where $P$ is the observed pulsar period, $V_{\rm \perp}^2/cD$ is the Shklovskii effect, $V_{\rm \perp}$ is the transverse velocity, $D$ is the distance to the pulsar, and $c$ is the speed of light. We assume the Galaxy has a flat rotation curve in the region of interest, with the Galactocentric distance of the Sun $R_{\rm 0}=8.0$~kpc and rotation velocity $\Theta=220\rm~km~s^{-1}$ \citep{rei93}. A straightforward geometric analysis gives the line-of-sight acceleration between the Sun and a pulsar at distance $D$ and Galactic {\bf coordinates} $l$ and $b$ as follows:
\begin{equation}
a_{\rm G}=-\cos b \left(\frac{\Theta_{\rm 0}}{R_{\rm 0}}\right) \left(\cos l + \frac{\beta}{\sin ^2 l+\beta^2}\right),
\end{equation}
where $\beta=(D/R_{\rm 0}) \cos b -\cos l$ \citep{dt91}. In the Shklovskii effect calculation, we use $V_{\rm \perp}=85\rm~km~s^{-1}$ which is the average velocity of MSPs \citep{tsb99}, as the velocity of PSR~J1853$-$0842A was not be fitted successfully with our observation data. If PSR~J1853$-$0842A lies outside the NGC~6712 at a distance of 4.4~kpc (the average DM-based distances list in Table~\ref{table:timingp}), the $a_{\rm G}/(cP)$ and $ V_{\rm \perp}^2 P/(cD)$ is $3.81\times10^{-22}$ and $3.42\times10^{-22}\rm~s~s^{-1}$, respectively. Taking $\dot P_{\rm int}>0$ and $a_{\rm L}/c P=0$ into consideration, it is impossible to obtained $\dot P_{\rm obs}=-2.39\times10^{-21}$ as all the terms on the right of Eq.~\ref{eq:pdot} are no less than 0. On the other hand, if PSR~J1853$-$0842A is located in NGC~6712 at the distance of 6.9~kpc, $a_{\rm G}/(cP)$ and $V_{\rm \perp}^2 P/(cD)$ is $2.43\times10^{-22}$ and $-9.35\times10^{-22}\rm~s~s^{-1}$ respectively with the same assumption $V_{\rm \perp}=85\rm~km~s^{-1}$. Compared the core position of NGC~6712 and PSR~J1853$-$0842A listed in Table~\ref{table:ngc6712} and ~\ref{table:timingp}, it is clear that the pulsar's projected distance
from the center of the GC ($R_{\rm psr}$ ) is less than two times of its core radius ($R_{\rm c}$). Under this condition, the maximum accelerations effect of GC can be estimated within 10\% of accuracy with the equations:
\begin{equation}
\frac{a_{\rm L, max}}{c}\approx\frac{3 \sigma_{\rm v}^2}{2 c (R_{\rm c}^2+R_{\rm psr}^2)^{1/2}}
\end{equation}
where $\sigma_{v}$ is the central velocity dispersion \citep{phi93}. For PSR~J1853$-$0842A, the $a_{\rm L, max}/c P$ is about $1.28\times10^{-20}\rm~s~s^{-1}$. Though we do not know three-dimensional position of PSR~J1853$-$0842A in NGC~6712 at present, its real $a_{\rm L}/c P$ is a certain value in the range of $-1.28\times10^{-20}$ to $1.28\times10^{-20}\rm~s~s^{-1}$. Accordingly, we give an upper limit of its intrinsic spin-down rate $\dot P_{\rm int, max}=1.11\times10^{-20}\rm~s~s^{-1}$. It is a reasonable value judging the $\dot P$ measure results of some other BWs. Taking PSRs~B1957+20 and J2051$-$0827 for example, their $\dot P$ is $2.7\times10^{-20}$ and $1.27\times10^{-20}\rm~s~s^{-1}$, respectively \citep{aft94, svf16}.

\begin{figure}
\centering
\includegraphics[width=0.49\textwidth]{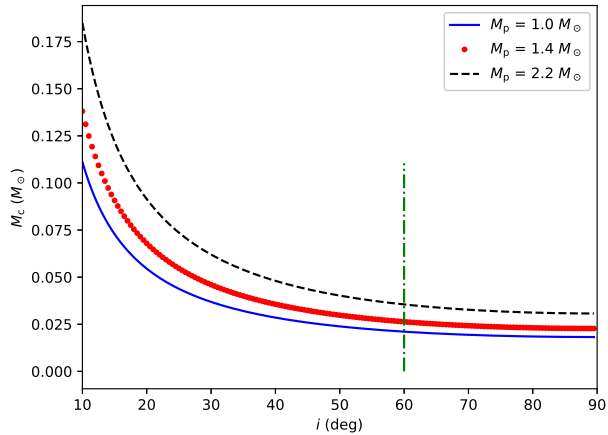}
\caption{The $M_{\rm c}$ as a function of $i$ and $M_{\rm p}$. The case for the assumption
$M_{\rm p} = 1.0~M_{\rm \odot}$, $M_{\rm p} = 1.4~M_{\rm \odot}$, and $M_{\rm p} = 2.2~M_{\rm \odot}$
is shown with solid, dotted and dash curve, respectively. The dash-dot vertical line stands for
$i = 60^{\circ}$}.
\label{fig:massincli}
\end{figure}

Using the orbital period $P_{\rm b}$ and the projected semi-major axis $x_{\rm p}$ of the pulsar orbit, the mass function
\begin{equation}
f(M_{\rm p}, M_{\rm c})=\frac{4\pi ^2}{G}\frac{x_{\rm p}^3}{P_{\rm b}^2}=\frac{(M_{\rm c} \sin i)^3}{(M_{\rm p}+M_{\rm c})^2},
\end{equation}
where $G$ is the gravitational constant, $M_{\rm p}$ is the mass of the pulsar, $M_{\rm c}$ is the mass of companion star, and $i$ is the inclination of the binary orbit. Though PSR~J1853$-$0842A is a short $P_{\rm b}$ binary, it is very hard to determine the mass of each star by measuring post-Keplerian parameters with pulsar timing because of its weak general relativity effects. As the $M_{\rm p}$ and $i$ of this binary can not be measured with our observation data, we estimate $M_{\rm c}$ under some assumptions. According to previous studies, the mass of a radio pulsar predominantly falls a remarkably narrow mass range $1.38^{-0.06}_{+0.10}M_{\rm \odot}$ \citep{tc99}. The canonical value of a pulsar 1.4~$M_{\rm \odot}$ is usually used in most cases. But, there are some exceptions, such as PSR~J1918-0642 with extremely low mass of 1.18~$M_{\rm \odot}$ \citep{fpe16} and PSR~J0740+6620 with extremely high mass of 2.14~$M_{\rm \odot}$ \citep{cfr20}. Taking these into consideration, further limitations on the $M_{\rm c}$ are given by assuming the $M_{\rm p}$ to be 1.0, 1.4, and 2.2~$M_{\rm \odot}$, respectively. The curves of $M_{\rm c}$ changing with $i$ and $M_{\rm p}$ are presented in Fig.~\ref{fig:massincli}. As the eclipse takes place in PSR~J1853$-$0842A, its inclination angle $i$ must be greater than $60^{\circ}$. We calculate $M_{\rm c, 60^\circ}$ for the different assumption of the $M_{\rm p}$ and listed related results in the 2nd column of Table~\ref{table:mcrerl}. Lower limits of the companion mass ($M_{\rm c, 90^\circ}$) are obtained by assuming an edge-on orbit ($i = 90^{\circ}$). The corresponding results of $M_{\rm c, 90^\circ}$
values are present in the 3rd column of Table~\ref{table:mcrerl}. The Roche lobe for the companion star of PSR~J1853$-$0842A is calculated by the equation:
\begin{equation}
R_{\rm L}=\frac{0.49 a q^{2/3}}{0.6 q^{2/3}+\ln(1+q^{1/3})},
\end{equation}
where $q=M_{\rm c}/M_{\rm p}$ and $a$ is the separation between the pulsar and its companion \citep{egg83}. The related results for different $M_{\rm p}$ and $i$ are present in the 4th (for $i = 60^\circ$) and 5th (for $i = 90^\circ$) column of Table~\ref{table:mcrerl}.

\begin{table}[htbp]
\centering
\caption{The results of $M_{\rm c}$, $R_{\rm L}$, $R_{\rm e}$ for different $i$ and $M_{\rm p}$}
\label{table:mcrerl}
\begin{tabular}{|c|c c| c c| c c|}
\hline
{$M_{\rm p}$} & \multicolumn{2}{c|}{$M_{\rm c}$~($M_{\rm \odot}$)} &\multicolumn{2}{c|}{$R_{\rm L}$ ~($R_{\rm \odot}$)} & \multicolumn{2}{c|}{$R_{\rm e}$~($R_{\rm \odot})$}  \\
\cline{2-7}
($M_{\rm \odot}$) & $M_{\rm c,60^{\circ}}$ & $M_{\rm c,90^{\circ}}$ & $R_{\rm L,60^{\circ}}$ & $R_{\rm L,90^{\circ}}$ & $R_{\rm e,60^{\circ}}$  & $R_{\rm e,90^{\circ}}$ \\
\hline
1.0 & 0.021& 0.018  &0.140&0.134 & 0.381 & 0.330 \\
1.4 & 0.026& 0.023  &0.152&0.145 & 0.427 & 0.370 \\
2.2 & 0.036& 0.031  &0.168&0.160 & 0.497 &  0.430 \\
\hline
\end{tabular}
\end{table}

About half of known BW pulsars show eclipse phenomena in which the radio pulses are completely blocked by companion stars \citep{goc19}. By comparison, the pulse signals of PSR~J1853$-$0842A are affected by extra time delay, but not absolutely blocked during eclipse phase. Base on the extra time delay lasted for about 10\% of its orbit at 1.25~GHz, we estimate the radius of ionized material surrounding its companion star ($R_{\rm e}$) for different orbital inclination angle and give the related results in the 6th (for $i = 60^\circ$) and 7th (for $i = 90^\circ$) column of Table~\ref{table:mcrerl}. It is clear that the $R_{\rm L}$ is smaller than $R_{\rm e}$, which indicates that ionized material fully fills the Roche lobe of the companion star. So, the outer material is being blown off the companion by the pulsar. As what is shown in this table, the $M_{\rm c}$, $R_{\rm L}$ and $R_{\rm e}$ is within a factor of 2.0, 1.3 and 1.5 of the minimum values, respectively. So, we will use the values corresponding to the case of canonical pulsar mass $M_{\rm p}=1.4~M_{\rm \odot}$ and $i=90^\circ$ in the following estimation about the properties of eclipse material.

Using the extra delays shown in the timing residual plot in Fig.~\ref{fig:toaplot}, we obtain the corresponding maximum excess dispersion measure ($\Delta_{\rm DM}$), which was about $0.028\pm0.001$, $0.036\pm0.001$, and $0.030\pm0.002$~$\rm pc~cm^{-3}$, respectively. The column density of electron ($N_{\rm e, max}$) in the eclipse material was about $9.69\times10^{16}\rm~cm^{-2}$ estimated with the average $\Delta_{\rm DM}$, which is about 5.7 times larger than that of BW PSR~J0023-7203J ($N_{\rm e, max}=1.7\times10^{16}\rm~cm^{-2}$). The radiations of PSR~J0023-7203J also passed the material around companion at both 660 and 1400~MHz with extra time delays \citep{fck03}. The radiations of several BWs are blocked at low radio frequencies, but get passed at higher frequencies with extra time delays. PSR~J1544+4937 was found to be eclipsing for 13\% of its orbit at 322~MHz, whereas the pulsar was detected throughout the low-frequency eclipsing phase at 607~MHz, which was affected by the material around its companion with $N_{\rm e, max}=8\times10^{16}\rm~cm^{-2}$ \citep{brr13}. The material with $N_{\rm e, max}\sim10^{17}\rm~cm^{-2}$ surrounding BW PSR J2051$-$0827's companion was also found to be opaque for radiations at 436 and 660~MHz, but transparent at 1.4~GHz \citep{sbl96}. By comparison, the $N_{\rm e, max}$ near the superior conjunction of BW PSR~J2055+3829 was no less than $10^{17}\rm~cm^{-2}$, which was larger than that of BWs mentioned above. And, its radiation was found to be blocked at 1.4~GHz in eclipse phase \citep{goc19}. Judging from the information mentioned above, we infer that radiation PSR~J1853$-$0842A has possibilities to be blocked at lower radio frequencies.

\begin{figure}
\centering
\includegraphics[width=0.50\textwidth]{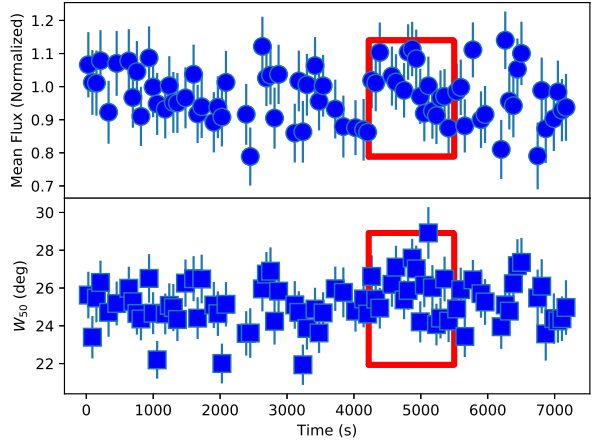}
\caption{The $W_{\rm 50}$ and normalized flux density changed with time for observation on MJD~58685. The eclipse time range is labeled with the red box.}
\label{fig:w50flux}
\end{figure}

Graduate flux density decreases (and increase) were detected in PSRs J2051$-$0827 and J2055+3829 before (and after) their radiations were completely eclipsed at the corresponding frequency \citep{sbl96, goc19}. For PSR~J1853$-$0842A, we also analyzed on its pulse flux density and shape changes with time to investigate how they changed in the eclipse phase. Considering the sensitivity and the time resolution, the $W_{\rm 50}$ were used in the pulse shape analysis, and the mean flux normalized with the peak flux density of integrated profile were used to seek its flux variations. Fig~\ref{fig:w50flux} shows how these two parameters changed in observation on MJD~58685, which spanned longest time and covered the full eclipse phase. No clear change in flux density and $W_{\rm 50}$ were found in the eclipse phase compared with normal state.

Base on the eclipse properties of PSR~J1853$-$0842A, we try to give some constrains on the accretion process of this system. From the results listed in Table~\ref{table:mcrerl}, it is clear that $R_{\rm e}$ does not change a lot for different assumptions of $i$ and $M_{\rm p}$. Considering only upper limit of $\dot P_{\rm int, max}$ obtained and eclipse-to-eclipse variability of extra DM delays, we only give rough estimations on the accretion with the average volume density of electron $n_{\rm e}=1.88\times10^6\rm~cm^{-3}$ by assuming $M_{\rm p}=1.4M_{\rm \odot}$ and $i=90^\circ$. The energy density of an isotropic pulsar wind at the distance of the companion is given by $U_{\rm E}=\dot E/(4 \pi c a^2)$, where $\dot E=4\pi^2 I \dot P P^{-3}$ is the spin-down power of the pulsar, $I$ is the moment of inertia and $a$ is the distance to the companion. For the canonical 1.4~$M_{\rm \odot}$ neutron star of radius 10~km and moment of inertia $I = 10^{45}~\rm g~cm^2$, the maximum $\dot E$ of PSR~J1853$-$0842A is about $4.42\times10^{34}\rm~erg~s^{-1}$. If the pulsar wind energy flow is converted into a mass outflow of kinetic energy density with an efficiency factor $\epsilon$, the outflow velocity of ablated material from the companion star,
\begin{equation}
V_{\rm W} \approx \left(\frac{2 \epsilon U_{\rm E}}{n_{\rm e} m_{\rm p}}\right)^{1/2},
\end{equation}
where $n_{\rm e}$ and $m_{\rm p}$ is the electron volume density and proton mass, respectively \citep{tbe94}. By assuming $\epsilon = 1$, the upper limit of mass loss rate from the companion  $\dot M_{\rm C}\simeq \pi R_{\rm e}^2 m_{\rm p} n_{\rm e} V_{\rm W}$. For PSR~J1853$-$0842A, the corresponding upper limit of $\dot M_{\rm C}$ should be about $3.05\times10^{-13}~M_{\rm \odot}\rm~yr^{-1}$. By comparison, the $\dot M_{\rm C}$ of PSR~J1810+1744 is $6\times10^{-13}~M_{\rm \odot}\rm~yr^{-1}$ and $1\times10^{-12}~M_{\rm \odot}\rm~yr^{-1}$ obtained block information at 149~MHz and 345~MHz, respectively \citep{pbc18}. The inferred accretion rates of these two BWs are comparable.

Overall, the newly-discovered BW pulsar J1853$-$0842A is most probably in NGC~6712. During the eclipse phase, its signals showed eclipse-to-eclipse varying extra time delays but were not blocked. The average $N_{\rm e, max}$ of material around J1853$-$0842A's companion is about $9.69\times10^{16}\rm~cm^{-2}$, which is a medium value compared with BWs that show only extra time delays and absolutely-blocked phenomena in the eclipse phase. Its inferred upper limit of $\dot M_{\rm C}$ is about $3.05\times10^{-13}~M_{\rm \odot}~\rm yr^{-1}$ is also comparable with some other BWs. We {\bf predict} that the radiation of PSR~J1853$-$0842A has the possibility to be blocked at lower radio frequencies, {\bf and will test this hypothesis with future observations.}

\section*{acknowledgments}
We would like to express our appreciations to the anonymous reviewer and Professor R.~N.~Manchester for their good suggestions. This work was supported in part by the National Natural Science Foundation of China (Grant Nos. U2031119, U1631122, 12041301, 11633007, and 11403073), the Strategic Priority Research Program of the Chinese Academy of Sciences (XDB23010200), the National Key R\&D Program of China (2018YFA0404602) and the Knowledge Innovation Program of the Chinese Academy of Sciences (KJCX1-YW-18). Five-hundred-meter Aperture Spherical radio Telescope (FAST) is a National Major Scientific Project built by the Chinese Academy of Sciences. Funding for the project has been provided by the National Development and Reform Commission. FAST is operated and managed by the National Astronomical Observatories, Chinese Academy of Sciences.


\end{document}